\documentclass[12pt]{article}
\usepackage[sort&compress]{natbib}
\bibpunct{[}{]}{,}{n}{}{}

\begin{document}
\sloppy

\title{Asymptotic behavior of Wilson loops from Schr\"{o}dinger equation on the gauge group}
\author{P. V. Buividovich$^{1}$, V. I. Kuvshinov$^{2}$, \\ 
\small{(1) Belarusian State University, Belarus,}\\ \small{220080 Minsk, F. Skoriny av. 4,}\\ \small{\emph{e-mail: buividovich@tut.by} (corresponding author)} \\ \small{(2) JIPNR, National Academy of Science, Belarus,}\\ \small{220109 Minsk, Acad. Krasin str. 99,}\\\small{ \emph{e-mail: v.kuvshinov@sosny.bas-net.by}}}
\date{November 10, 2005}
\maketitle

\begin{abstract}
Probability distribution of non-Abelian parallel transporters on the group manifold and the corresponding amplitude are investigated for quantum Yang-Mills fields. It is shown that when the Wilson area law and the Casimir scaling hold for the quantum gauge field, this amplitude can be obtained as the solution of the free Schr\"{o}dinger equation on the gauge group. Solution of this equation is written in terms of the path integral and the corresponding action term is interpreted geometrically. We also note that the partition function of two-dimensional pure Yang-Mills theory living on the surface spanned on the loop solves the obtained equation.
\end{abstract}

{\footnotesize PACS numbers: 11.15.Me; 12.38.Aw; 12.40.Ee}

\section{Introduction}

 It is known that Yang-Mills theory can be formulated in terms of loop variables \cite{PolyakovGaugeStrings, Polyakov:79, Polyakov:80, Migdal:81}. In this approach basic variables describing gauge fields are not the gauge field vector $\hat{A}_{\mu}$ or the curvature tensor $\hat{F}_{\mu \nu}$, but rather parallel transport operators (non-Abelian phase factors) associated with loops in physical space \cite{Polyakov:79, Polyakov:80, Migdal:81} or their traces (Wilson loops). This formalism allows one to consider the dynamics of Yang-Mills fields as dynamics of chiral fields on loop space and is particularly useful in investigating the relation between gauge theories and string theories \cite{PolyakovGaugeStrings, Polyakov:79}. Asymptotic behavior of loop variables at large distances is known to be closely connected with the confining properties of the theory \cite{Wilson:74, Ambjorn:80, Dosch:02, Simonov:96}. In the conventional picture of confinement colour charges are connected by QCD string, which corresponds to the well-known Wilson area law for the Wilson loop \cite{Wilson:74}. 

 On the other hand Wilson loops are directly measured in various lattice simulations. Besides Wilson area law, results of simulations of four-dimensional pure Yang-Mills theory \cite{Bali:00, Deldar:00} confirmed the Casimir scaling phenomenon, predicted by many models of nonperturbative QCD vacuum \cite{Dosch:02, Ambjorn:80, Simonov:96, Steffen:03}.

 These most general results of lattice simulations can be used to reveal some aspects of loop dynamics. In this paper we investigate the probability distribution of parallel transporters on the group manifold and the corresponding amplitude, which is the quantum counterpart of classical loop variables. It is shown that when the Wilson area law \cite{Wilson:74} and Casimir scaling \cite{Dosch:02, Simonov:96} hold for quantum gauge field, this amplitude can be obtained as the solution of the free Schr\"{o}dinger equation on the group manifold. Solutions of this equation are then expressed in terms of path integral. A nontrivial conclusion is that two-dimensional pure Yang-Mills theory living on the surface spanned on the loop also solves this equation. This fact may provide some hints on the correct string theory which solves quantum loop equations \cite{PolyakovGaugeStrings, Polyakov:80}. It is also in agreement with the dimensional reduction scenario in lattice gauge theory \cite{Makeenko:84}. We also discuss possible modification of the obtained equation when Casimir scaling is violated.

\section{Distribution of parallel transporters on the group manifold}

We consider gauge fields $\hat{A}_{\mu} (x)$ belonging to the Lie algebra of some simple compact group $G$. To any smooth closed path $\gamma$ we can attribute parallel transport operator defined as the path-ordered exponent:
\begin{equation}
\label{ParallelTransportOperatorDefinition}
    \hat{U} \left( \gamma \right) = \mathcal{P} \exp \left( i \int \limits_{\gamma} dx^{\mu} \hat{A}_{\mu} \right)
\end{equation}   
Parallel transport operator is the element of some representation of the group $G$, therefore to each path $\gamma$ corresponds some group element $g(\gamma)$. For the sake of simplicity here we will consider only topologically trivial loops, without intersections or twists. In order to evaluate the path-ordered exponent in (\ref{ParallelTransportOperatorDefinition}) one should introduce some initial point $x_{0}$ on the loop $\gamma$. Parallel transport operator transforms under gauge transformations and shifts of the initial point $x_{0} \rightarrow x_{0}'$ in the following way:
\begin{equation}
\label{ParallelTransporterGaugeTransforms}
\begin{array}{l}
    \hat{U} \left( \gamma \right) \rightarrow \hat{T} (x_{0}) \hat{U} \left( \gamma \right) \hat{T}^{\dag} (x_{0}) \\
\hat{U} \left( \gamma \right) \rightarrow  
\mathcal{P} \exp \left( i \int \limits_{\gamma; x_{0}}^{x_{0}'} dx^{\mu} \hat{A}_{\mu} \right)
\hat{U} \left( \gamma \right) 
\mathcal{P} \exp \left( i \int \limits_{\gamma; x_{0}'}^{x_{0}} dx^{\mu} \hat{A}_{\mu} \right)
\end{array}
\end{equation}

 In quantum theory one can not establish any deterministic correspondence between loops in physical space and the elements of the gauge group $G$, but one can calculate the amplitude which will determine the probability $dP(g)$ for the group element $g(\gamma)$ to be within an infinitesimal volume $d \mu(g)$ on the group manifold. We introduce this amplitude $\psi(g ; \: \gamma)$ in the following way:
\begin{equation}
\label{AmplitudeDefinition}
    \psi(g ; \: \gamma) = \langle \: \delta \left(g,g(\gamma) \right) \: \rangle
\end{equation} 
where by $\langle \: \ldots \: \rangle$ we denote vacuum expectation value (we work in pure euclidean Yang-Mills theory without quarks) and $\delta \left(f,g\right)$ is the delta-function on the group manifold defined by the following relation:
\begin{equation}
\label{GroupDeltaFunctionDefinition}
 \int \limits_{G} d \mu (g) \phi(g) \delta \left(g,f \right) = \phi(f)
\end{equation}
for an arbitrary function $\phi(g)$ on the group manifold. Here $d \mu (g)$ is the Haar measure on the group $G$ normalized as $\int \limits_{G} d \mu(g) = 1$. The probability $dP(g)$ is then obtained as:
\begin{equation}
\label{ProbabilityDistributionOnTheGroup}
    dP(g) = |\psi(g ; \: \gamma)|^{2}  d \mu (g)
\end{equation}

Gauge invariance of the theory implies that the amplitude $\psi(g ; \: \gamma)$ should be also invariant under gauge transformations and in addition it should not depend on the choice of the initial point $x_{0}$ on the loop. It follows then from the transformation laws (\ref{ParallelTransporterGaugeTransforms}) and from the fact that $d \mu (g) = d \mu (f^{-1} g \: f)$ for any group element $f$ that the amplitude $\psi(g ; \: \gamma)$ should actually be the function on the group classes \cite{BarutRonczkaGroupRepresentation, LittlewoodGroupCharacters}, that is:
\begin{equation}
\label{DistributionSymmetry}
    \psi(g ; \: \gamma) = \psi(f^{-1} g \: f; \: \gamma)
\end{equation}
for any element $f$ of the gauge group $G$.

It is known from the theory of harmonic analysis on compact groups that the characters of irreducible unitary representations of simple compact group build full orthonormal basis in the Hilbert space of functions on the group classes \cite{BarutRonczkaGroupRepresentation, LittlewoodGroupCharacters}, therefore the function $\psi(g ; \: \gamma)$ can be decomposed as:
\begin{equation}
\label{ProbabilityDistributionDecomposition}
 \psi(g; \: \gamma) = \sum \limits_{k} \bar{\chi}_{k}(g) \psi_{k} (\gamma)
\end{equation}
where the subscript $k$ labels irreducible unitary representations of the group $G$, including the trivial one, $\chi_{k}(g) = {\rm Tr} \: \: \hat{T}^{(k)}(g)$ are the group characters and $\hat{T}^{(k)}(g)$ is the matrix in the $k$-th representation which corresponds to the element $g$. Using the orthogonality of group characters $ \int \limits_{G} d \mu (g) \chi_{k}(g) \bar{\chi}_{l}(g) = \delta_{kl}$ one obtains for the decomposition coefficients $\psi_{k} (\gamma)$ in (\ref{ProbabilityDistributionDecomposition}):
\begin{equation}
\label{DecompositionCoefficients}
\begin{array}{l}
 \psi_{k} (\gamma) = \int \limits_{G} d \mu(g) \chi_{k}(g) \psi(g; \: \gamma) =
 \int \limits_{G} d \mu(g) \chi_{k}(g) \langle \: \delta \left(g,g(\gamma) \right) \: \rangle = \\ = 
 \langle \: \int \limits_{G} d \mu(g) \chi_{k}(g)  \delta \left(g,g(\gamma) \right) \: \rangle = 
 \langle \: \chi_{k}\left(g (\gamma) \right) \: \rangle
 \end{array}
\end{equation}  

Expression (\ref{DecompositionCoefficients}) shows that the coefficients $\psi_{k} (\gamma)$ are the vacuum expectation values of the traces of parallel transporters calculated in the $k$-th representation of the group $G$:
\begin{equation}
\label{DecompositionCoefficientsViaWilsonLoops}
 \psi_{k}(\gamma) = \langle \: {\rm Tr} \: \mathcal{P} \exp \left( i \int \limits_{\gamma} dx^{\mu} \hat{A}^{(k)}_{\mu} \right)  \: \rangle
\end{equation}
It is convenient to express the quantities $\psi_{k} (\gamma)$ in terms of the Wilson loops $W_{k} (\gamma)$ which are conventionally used in nonperturbative QCD \cite{Wilson:74, Dosch:02, Simonov:96}:
\begin{equation}
\label{WilsonLoopDefinition}
    W_{k} (\gamma) = d_{k}^{-1} \langle \: {\rm Tr} \: \mathcal{P} \exp \left( i \int \limits_{\gamma} dx^{\mu} \hat{A}^{(k)}_{\mu} \right)  \: \rangle
\end{equation}
where $d_{k}$ is the dimensionality of $k$-th representation. Using Wilson loops one can rewrite decomposition (\ref{ProbabilityDistributionDecomposition}) as:
\begin{equation}
\label{ProbabilityDistributionDecompositionWithWilsonLoops}
    \psi(g; \: \gamma) = \sum \limits_{k} \bar{\chi}_{k}(g) d_{k} W_{k} (\gamma)
\end{equation}
Such decomposition clarifies the physical meaning of the amplitude $\psi(g; \: \gamma)$: as the Wilson loop calculated in $k$-th representation characterizes interaction of static colour charges in this representation \cite{Dosch:02, Ambjorn:80, Simonov:96}, the function $\psi(g; \: \gamma)$ describes simultaneously interaction between colour charges in all representations.

\section{Wilson area law and Casimir scaling}

 In the previous section we have managed to express the amplitude $\psi(g; \: \gamma)$ in terms of Wilson loops, which are physically observable, gauge-invariant quantities. The properties of Wilson loops are known from lattice simulations and from predictions of different phenomenological models \cite{Wilson:74, Ambjorn:80, Dosch:02, Simonov:96}. The most general conclusion which can be drawn basing on the lattice data obtained for Yang-Mills theory without quarks is the validity of Wilson area law and Casimir scaling. In general both Casimir scaling and Wilson area law can be violated at large distances due to screening effects, especially in the presence of dynamical fermions, but in lattice simulations of pure $SU(3)$ Yang-Mills theory no such violation was found up to the largest available distances \cite{Bali:00, Deldar:00}. The level of violation did not exceed few percents in the results \cite{Bali:00} and $5 - 15 \%$ in the measurements of \cite{Deldar:00}. Extrapolation to the continuum limit performed in \cite{Bali:00} indicated no  violation of Casimir scaling except statistical errors, of the order of $1 \%$. Casimir scaling holds also for small distances, as is indicated by perturbative calculations with two loops \cite{Peter:97}. However some theoretical considerations predict that violation may occur \cite{Greensite:83, Ambjorn:80}. Thus we will restrict our analysis to the case of intermediate distances ($\sim 0.2 \ldots 1$ fm) where the validity of both Wilson area law and Casimir scaling is checked with good precision. 

 Wilson area law means that $W_{k} (\gamma) = \exp( - \sigma_{k} S_{\gamma})$, where $\sigma_{k}$ is the tension of QCD string between charges in $k$-th representation \cite{Ambjorn:80, Dosch:02, Simonov:96} and $S_{\gamma}$ is the minimal area of the surface spanned on the loop $\gamma$. Wilson area law implies linear interaction potential $V(R) = \sigma_{k} R$ between static colour charges \cite{Wilson:74}. If Casimir scaling holds, the tension $\sigma_{k}$ is proportional to the eigenvalue of quadratic Casimir operator in $k$-th representation: $\sigma_{k} = \sigma C_{k}$, where $\sigma$ is some constant. It can be shown using the field correlator method that $\sigma$ is proportional to the density of nonperturbative gluonic condensate $\sigma \sim \langle \: {\rm Tr} \: \hat{F}_{\mu \nu} \hat{F}^{\mu \nu} \: \rangle$ \cite{Dosch:02, Simonov:96}. Expansion (\ref{ProbabilityDistributionDecompositionWithWilsonLoops}) can be now rewritten as:
\begin{equation}
\label{ProbabilityDistributionAfterAssumptions}
    \psi(g; \: \gamma) = \sum \limits_{k} \bar{\chi}_{k}(g) d_{k} \exp{ \left( - \sigma C_{k} S_{\gamma} \right)}
\end{equation}
After differentiating (\ref{ProbabilityDistributionAfterAssumptions}) over $S_{\gamma}$ one obtains:
\begin{equation}
\label{ProbabilityDistributionDerivative}
    \frac{d}{d S_{\gamma}} \: \psi(g; \: \gamma) = \sum \limits_{k} - \sigma C_{k} \bar{\chi}_{k}(g) \: d_{k} \exp{ \left( - \sigma C_{k} S_{\gamma} \right)}
\end{equation} 

 In order to rewrite (\ref{ProbabilityDistributionAfterAssumptions}) and (\ref{ProbabilityDistributionDerivative}) in a compact beautiful form it is convenient to use some geometric constructions on the group manifold. Suppose that the elements of the group $G$ are parameterized by the coordinates $\lambda^{\alpha}$. The metrics on the group manifold (Killing form) is fixed up to a constant multiplier by requiring the infinitesimal distance to be invariant under left and right group multiplications \cite{DeWittGroupsAndFields, BarutRonczkaGroupRepresentation, LittlewoodGroupCharacters}. The metric tensor of the form:
\begin{equation}
\label{GroupMetricFixed}
    g_{\alpha \beta} (g) = {\rm Tr} \: \left( \frac{\partial}{\partial \lambda^{\alpha}} \hat{T} (g) \frac{\partial}{\partial \lambda^{\beta}}  \hat{T}^{\dag } (g) \right)
\end{equation}
with $\hat{T} (g)$ being the matrix of some irreducible unitary representation (for example, fundamental) of the group $G$, satisfies this condition \cite{BarutRonczkaGroupRepresentation, LittlewoodGroupCharacters}. The Haar measure is $d \mu (g) = \sqrt{ \det {g_{\alpha \beta}}} \prod \limits_{\delta} d \lambda^{\delta}$. Group structure of the manifold also induces the connection on it, which can be used to build the covariant derivative $\nabla_{\alpha}$ \cite{DeWittGroupsAndFields}. An important property of $\nabla_{\alpha}$ is that the curvature tensor associated with it is equal to zero, thus group manifold has zero curvature but nonzero torsion $S^{\gamma}_{\alpha \beta}$: $[\nabla_{\alpha}, \nabla_{\beta}] = S^{\gamma}_{\alpha \beta} \nabla_{\gamma}$. An essential fact is that covariant derivatives acting on the Hilbert space of smooth functions on group manifold build a reducible infinite-dimensional representation of the Lie algebra of the group $G$ \cite{BarutRonczkaGroupRepresentation, DeWittGroupsAndFields} which contains all irreducible finite-dimensional representations. Group laplacian $\Delta$ is the quadratic Casimir operator of this reducible representation:
\begin{equation}
\label{GroupLaplacian}
    \Delta = g^{\alpha \beta} \nabla_{\alpha} \nabla_{\beta}
\end{equation}
Group laplacian was introduced by Berezin in  \cite{Berezin:62}. Subspaces of irreducible representations are the eigenspaces of $\Delta$ corresponding to eigenvalues $ - C_{k}$. For instance, for the characters of irreducible representations we have  \cite{BarutRonczkaGroupRepresentation, LittlewoodGroupCharacters, Berezin:62}:
\begin{equation}
\label{CharactersLaplacian}
 \Delta \chi_{k}(g) = - C_{k} \chi_{k}(g)
\end{equation}
The equation (\ref{ProbabilityDistributionDerivative}) can be now rewritten as:
\begin{equation}
\label{FinalDiffusionEquationDerivation}
    \frac{d}{d S_{\gamma}} \: \psi(g; \: \gamma) = \sum \limits_{k} \sigma \Delta \bar{\chi}_{k}(g) \: d_{k} \exp{ \left( - \sigma C_{k} S_{\gamma} \right)} = \sigma \Delta \psi(g; \: \gamma)
\end{equation}
Thus the final result for the amplitude $\psi(g; \: \gamma)$ has the form of the Schr\"{o}dinger equation on the group manifold \cite{KleinertPathIntegrals, Kleinert:90} with $S_{\gamma}$ playing the role of the euclidean time:
\begin{equation}
\label{FinalDiffusionEquation}
    \frac{d}{d S_{\gamma}} \: \psi(g; \: \gamma) = \sigma \Delta \psi(g; \: \gamma)
\end{equation}
Initial condition for the equation (\ref{FinalDiffusionEquation}) is $\psi(g; \: \gamma)|_{S_{\gamma}=0} = \delta(g,1)$,  which means that for infinitely small loops parallel transport operator is always equal to identity. Equation (\ref{FinalDiffusionEquation}) is also known as the diffusion equation on the group manifold \cite{Varopoulos:94,  Varopoulos:96}.

 Finally we would like to discuss how possible violation of Casimir scaling due to screening effects can be taken into account. From general analysis based on the cumulant expansion of Wilson loops \cite{Dosch:02, Simonov:96} it is known that Casimir scaling is violated by contributions to the string tension which are proportional to higher-order Casimir operators. As covariant derivatives build a representation of the Lie algebra of the gauge group, in order to reproduce higher-order Casimir operators one should act on $\psi(g; \gamma)$ with covariant derivatives more than twice, which leads to the following generalization of the equation (\ref{FinalDiffusionEquation}):
\begin{equation}
\label{KramersMoyal}
 \frac{d}{d S_{\gamma}} \: \psi(g; \: \gamma) = \sum \limits_{l} \sigma^{\alpha_{1} \ldots \alpha_{l}}(\gamma) 
 \nabla_{\alpha_{1}} \ldots \nabla_{\alpha_{l}}
  \psi(g; \: \gamma)
\end{equation}
Gauge invariance imposes rather strict constraints on the tensor structure of the coefficients $\sigma^{\alpha_{1} \ldots \alpha_{l}}(\gamma)$. Violation of the Wilson area law can be implemented in the functional dependence of $\sigma^{\alpha_{1} \ldots \alpha_{l}}(\gamma)$ on $\gamma$. Such equation is known in the theory of random walks and diffusions as the Kramers-Moyall expansion \cite{RiskenFokkerPlanck}. Application of this equation to the description of Wilson loops in QCD deserves separate discussion and will be investigated in subsequent publications. 

\section{Path integral solution and two-dimensional Yang-Mills theory}

Let us now consider the set of rectangular loops $\gamma_{R \times t}$ which stretch the time $t$ and the distance $R$. Assuming that $R$ is constant, one can rewrite (\ref{FinalDiffusionEquation}) as the Schr\"{o}dinger equation on the group manifold \cite{KleinertPathIntegrals} in euclidean time $t$:
\begin{equation}
\label{SchrodingerEuclide}
    \frac{d}{d t} \: \psi(g, t) = \sigma R \: \Delta \psi(g, t)
\end{equation}
where by $\psi(g, t)$ we denoted the amplitude for the rectangular loop of the size $R \times t$. Typically such rectangular Wilson loop is used to describe interaction of two static colour charges separated by distance $R$. 

The equation (\ref{SchrodingerEuclide}) is a Schr\"{o}dinger equation for a point on the group manifold \cite{Kleinert:90, KleinertPathIntegrals} with the following lagrangian:
\begin{equation}
\label{ActionTerm}
L = (4 \sigma R)^{-1} g_{\alpha \beta} \frac{d \lambda^{\alpha}}{d t} \frac{d \lambda^{\beta}}{d t} 
\end{equation}
Just as in the case of quantum mechanics in euclidean three-dimensional space, solutions of the equation (\ref{SchrodingerEuclide}) can be expressed in terms of path integrals \cite{Fiziev:96, Kleinert:90} with this lagrangian and the proper integration measure:
\begin{equation}
\label{ProbabilityDistrViaPathInt}
\psi(g, t) = \int \limits_{f(0) = 1}^{f(t) = g} \mathcal{D} \mu \left[ f(t) \right]
\exp \left(
 - \frac{1}{4 \sigma R} \int \limits_{0}^{t} d \tau \: g_{\alpha \beta}  (\chi) \frac{d \lambda^{\alpha}}{d \tau} \frac{d \lambda^{\beta}}{d \tau}
\right)
\end{equation}
In this expression integration is performed over all paths on the group manifold which begin at the time $\tau=0$ in the identity and end at the time $\tau = t$ in the point $g$. In (\ref{ProbabilityDistrViaPathInt}) $f(\tau)$ is the group element passed at time $\tau$, $\lambda^{\alpha} (\tau)$ are the coordinates which correspond to $f(\tau)$ and $\mathcal{D} \mu \left[ f(t) \right] = \prod \limits_{\tau} d \mu \left(f(\tau) \right)$ is the path-integral measure \cite{KleinertPathIntegrals}. Substituting (\ref{GroupMetricFixed}) in the lagrangian (\ref{ActionTerm}) we find the action in the path integral:
\begin{equation}
\label{ActionInTermsOfGroupElements}
    S \left[f(\tau)  \right] = (4 \sigma R)^{-1} \int \limits_{0}^{t} {\rm Tr} \: \left( d \hat{T} \left(g(\tau) \right) \frac{d \hat{T}^{\dag} \left(g(\tau) \right)}{d \tau} \right)
\end{equation}
where $g(\tau)$ is the group element corresponding to the loop $\gamma_{R \times \tau}$. Now remember that $\hat{T}\left(g(\tau) \right)$ is the parallel transporter along this loop. Writing the differentials in (\ref{ActionInTermsOfGroupElements}) as infinitesimal differences, we obtain for the action (\ref{ActionInTermsOfGroupElements}):
\begin{equation}
\label{ActionReWrittenViaWilsonLoops}
\begin{array}{l}
    S =  \int \limits_{0}^{t} (4 \sigma R \: d \tau)^{-1} {\rm Tr} \: \left(
  \left[ \hat{T} \left(g(\tau + d \tau) \right) - \hat{T} \left(g(\tau) \right) \right] 
  \left[\hat{T}^{\dag} \left(g(\tau + d \tau) \right) - \hat{T}^{\dag} \left(g(\tau) \right) \right]
 \right) = \\ =
  \int \limits_{0}^{t} (4 \sigma R \: d \tau)^{-1}
 {\rm Tr} \: \left(2 \hat{I} -  \hat{T} \left(g(\gamma_{R \times d \tau}) \right) - \hat{T}^{\dag} \left(g(\gamma_{R \times d \tau}) \right) \right) = \\ = {\rm Tr} \: \hat{I} \int \limits_{0}^{t} (2 \sigma R \: d \tau)^{-1}
  \left(1 - ( {\rm Tr} \: \hat{I})^{-1} {\rm Re \: Tr} \: \hat{U} (\gamma_{R \times d \tau}) \right) = \\ =  {\rm Tr} \: \hat{I} \int \limits_{0}^{t} (2 \sigma R \: d \tau)^{-1}
  \left(1 - {\rm Re} \: W (\gamma_{R \times d \tau}) \right)
\end{array} 
\end{equation}
where $\gamma_{R \times d \tau}$ is the rectangular loop of size $R \times d \tau$ situated in the point $\tau$ on the time axis (such loop can be though of as the difference of two loops $\gamma_{R \times (\tau + d \tau)}$ and $\gamma_{R \times \tau}$) and $\hat{U} (\gamma_{R \times d \tau}) = \mathcal{P} \exp \left( i \int \limits_{\gamma_{R \times d \tau}} dx^{\mu} \hat{A}^{(k)}_{\mu} \right)$. Thus we arrive at the geometric interpretation of the action term (\ref{ActionTerm}): it is expressed in terms of the Wilson loop for an infinitely thin contour of size $R \times d \tau$. One can expand $\hat{U} (\gamma_{R \times d \tau})$ in powers of loop area using the non-Abelian Stokes theorem \cite{Fishbane:81, Diosi:83}, which shows that the first nonvanishing term in $1 - {\rm Re} W_{k} (\gamma_{R \times d \tau})$ is proportional to the square of the loop area, therefore the action differential is of order $d \tau$, and the integral in (\ref{ActionReWrittenViaWilsonLoops}) is mathematically well-defined.

 Note that the action term in (\ref{ActionReWrittenViaWilsonLoops}) resembles the Wilson action for two-dimensional QCD. This resemblance is not accidental and provides another possibility to treat the equation (\ref{FinalDiffusionEquation}). Consider pure two-dimensional Yang-Mills theory with gauge group $G$ on the surface of the minimal area $S_{\gamma}$ spanned on the loop $\gamma$ and fix the boundary condition by assuming that the parallel transporter along the loop $\gamma$ (calculated using the connection of two-dimensional theory) is equal to some group element $g$. Such theory is exactly solvable, which was first demonstrated in the remarkable paper by Migdal \cite{Migdal:76}. For instance, partition function of the theory is:
\begin{equation}
\label{QCD2DPartition}
 \mathcal{Z}(g,S_{\gamma}) = \sum \limits_{k} d_{k} \chi_{k} (g) \exp (- \sigma_{2D} C_{k} S_{\gamma} )
\end{equation}
where $\sigma_{2D}$ is some constant which depends on the normalization of the coupling constant. By definition partition function is the sum over all field configurations of our two-dimensional theory which satisfy the boundary condition. The function $\mathcal{Z}(g,S_{\gamma})$ becomes the solution of the equation (\ref{FinalDiffusionEquation}) with the proper initial conditions $\mathcal{Z}(g,0) = \delta(g,1)$, if one assumes $\sigma_{2D} = \sigma$. Thus two-dimensional pure Yang-Mills theory living on the surface $S_{\gamma}$ satisfies the equation (\ref{FinalDiffusionEquation}). This could be expected, since both Casimir scaling and Wilson area law are exact for such theory. It is interesting whether the equation (\ref{KramersMoyal}) can also be solved by means of some path integral.

\section{Conclusions}

 In this work we have analyzed the behavior of a single loop variable in quantum gauge field. The amplitude which corresponds to the probability distribution of loop variables on the group manifold was expressed in terms of the Wilson loops in different representations of the gauge group. It was shown that the Wilson area law and the Casimir scaling can be united, leading to the free Schr\"{o}dinger equation on the group manifold with euclidean "time" being the area of the surface spanned on the loop. Possible screening effects can be taken into account if one uses a generalization of the simple diffusion equation (\ref{FinalDiffusionEquation}) known as the Kramers-Moyall expansion.

 The solution of the obtained equation was written in terms of the path-integral and the corresponding action term was analyzed. It turned out that the action differential can be linearly expressed in terms of an infinitely thin Wilson loop, which resembles the Wilson action for two-dimensional Yang-Mills theory. Next we demonstrated that pure two-dimensional Yang-Mills theory living on the surface $S_{\gamma}$ spanned on the loop $\gamma$ also solves the equation (\ref{FinalDiffusionEquation}). This points at some relation between two-dimensional and four-dimensional gauge theories, which is known in somewhat different context in lattice gauge theory as the dimensional reduction scenario \cite{Makeenko:84}. One can also surmise that two-dimensional gauge theory can provide some approximate solution of the quantum loop equations of four-dimensional theory \cite{PolyakovGaugeStrings, Migdal:81, Polyakov:79, Polyakov:80}, but this possibility require further investigations.

 As a further development of this result it can be useful to look at the joint probability distribution of two or more loop variables. The case of loops with self-intersections can be particularly interesting in connection with the quantum loop equations \cite{PolyakovGaugeStrings, Migdal:81, Polyakov:79, Polyakov:80}.


\end{document}